\documentclass[9pt]{article}
\usepackage{amsfonts}
\usepackage{pstricks}
\usepackage{pst-node}
\usepackage{epsfig}
\usepackage{epsfig}
\usepackage{graphicx}
\usepackage[font={footnotesize,it}]{caption}
\usepackage[latin5]{inputenc}
\usepackage[T1]{fontenc}
\usepackage{lipsum}
\usepackage{amsmath}
\usepackage{cite}
\begin{document}
	\def\ba{\begin{eqnarray}}
		\def\ea{\end{eqnarray}}
	\def\w{\wedge}
	\def\d{\mbox{d}}
	\def\D{\mbox{D}}
	\begin{titlepage}
		\title{Majorana Neutrinos in Sandwich Wave Spacetimes: Flavor Modulation and Helicity Transitions}
		\author{Yorgo Senikoglu \footnote{ysenikoglu@gsu.edu.tr}}
		\date{%
		 \small D\'{e}partement de Math\'{e}matiques, Universit\'{e} Galatasaray, 34349 Be\c{s}ikta\c{s}, \.{I}stanbul, Turkey\\
			\today}
		
		\maketitle
\begin{abstract}
	We present the propagation of massive Majorana neutrinos crossing an exact gravitational sandwich wave background. By employing a Takagi factorization to decouple the matrix field equations, We show that the localized pulse breaks the kinematic alignment of the mass eigenstates as they propagate through the wave zone, leaving a permanent residual phase shift across the transverse profile of the wavefront. The wave asymmetry couples directly to the Majorana spin tensor, forcing a transition from left-handed to right-handed configurations. Behind the wave tail, the residual drift parameters act to polarize the state's chirality, introducing a strong anisotropy across the transverse momentum plane. These results show that a passing gravitational wave leaves a modification on both the flavor and helicity of a neutrino beam, as a sign of the gravitational memory effect.
\end{abstract}

\end{titlepage}

\section{Introduction}

The behavior of neutrinos moving through intense gravitational environments sits at an important intersection connecting quantum field theory, particle astrophysics and general relativity. Since the first studies of neutrino fields in curved spacetimes \cite{brill-wheeler, wainwright, trim-wainwright}, it has been established that strong gravitational backgrounds can deeply change the kinematic and topological properties of matter fields. Following the direct observation of gravitational radiation from binary black hole mergers by the LIGO-Virgo collaborations \cite{abbott et al}, analyzing particle propagation through dynamic spacetimes has become a key element of multi-messenger astrophysics.

Historically, the coupling between neutrinos and gravitational radiation has been evaluated primarily through the in the weak-field limit of general relativity or within cosmological models tracking the impact of cosmic backgrounds \cite{Mangilli}. While these perturbative models provide valuable weak-field insights, they fail to capture the non-linear, non-perturbative physics that only emerges within exact solutions of the Einstein field equations. To probe the full, exact back-reaction and coupling between spacetime curvature and quantum fields, one must use exact plane wave spacetimes. These geometries were introduced by Brinkmann \cite{brinkmann}, Rosen \cite{rosen}, Einstein \cite{einstein-rosen}, and Peres \cite{peres}, and were cataloged by Ehlers, Kundt, Pirani, and Podolsky \cite{ehlers-kundt, pirani, griffiths-podolsky}.

The sandwich wave profile is a convenient choice for this framework. In this configuration, the spacetime curvature is exists inside a finite null-duration interval ($0 \leq u \leq u_0$), becoming asymptotically flat Minkowski domains on either side \cite{halilsoy, albadawi-halilsoy}. Analyzing interactions in these regions reveals unique energy amplification and particle creation signatures during wave collisions \cite{penrose, szekeres, griffiths0, bell-szekeres, abbasi1, abbasi2}. When analyzed in Rosen coordinates, a passing sandwich wave leaves behind a residual physical consequence known as the gravitational velocity memory effect \cite{ZHANG2017743, Maluf}. Consequently, while the Riemann curvature vanishes identically in the after region, the metric functions preserve a permanent shift with non-zero drift velocities.

Inside these wave geometries, the non-linear metric coefficients explicitly separate the transverse momentum components of the fermion field. Early calculations showed that exact plane-wave backgrounds alter the quantum states of propagating fields \cite{gibbons, griffiths1, griffiths2}. Recent studies have extended this to analyze Dirac particle scattering and spin-flavor oscillations in curved spacetimes \cite{dvornikov1, dvornikov2, bini-ferrari2, collas}. However, the situation changes for massive Majorana neutrinos. Because a linearly polarized plane wave breaks local isotropy, the spin connection directly couples the active left-handed and right-handed fields. While prior work successfully mapped standard neutrino fields in pure sandwich wave metrics \cite{dereli-senikoglu} and explored formal intrinsic coupling constraints \cite{dereli-tucker}, a full analytical description of Majorana flavor modulation combined with gravitationally induced helicity transitions remains open.

In this paper, we solve the exact propagation of a two-flavor massive Majorana neutrino system across a purely gravitational sandwich plane wave spacetime. By projecting the matrix Weyl equations onto a null tetrad, we obtain a system of differential equations governed by the metric functions F(u) and G(u). Using the Takagi\cite{Takagi} factorization, we decouple the mass eigenstates and find explicit analytical phase solutions integrated piecewise across each spacetime region.

Using this exact framework, we identify two physical phenomena that differ from standard linearized gravity results. First, the gravitational effect breaks the kinematic alignment ($p_\mu \neq k_\mu$), causing the neutrino beam to develop a transverse interference pattern with a permanent phase memory $\Delta\Phi_{\text{mem}}$. Second, the residual velocity memory of the wave mixes left and right-handed states. The resulting conversion rate shows a strong transverse momentum dependence, directly linking the gravitational memory to quantum observables.

The paper is organized as follows. In Section 2, we outline the geometry of the gravitational sandwich wave, explicitly defining the Brinkmann and Rosen coordinate representations and their junction conditions. In Section 3, we construct the coupled Majorana flavor equations, introduce the Takagi diagonalization method, and present the completely decoupled exact wave solutions. Section 4 provides a detailed analysis of the physical observables, tracking the dynamic flavor survival probabilities, the gravitationally induced helicity transitions, and the resulting transverse momentum anisotropies. Finally, we offer concluding remarks in Section 5.

\bigskip

\section{Sandwich Gravitational Plane Waves}
\noindent
To describe the spacetime, we adopt null coordinates aligned with the wave: $U$ represents the advanced null coordinate along the direction of propagation, while $X$ and $Y$ serve as the transverse spatial coordinates labeling the wave fronts.

Specifically:
\ba
U=\frac{z+t}{\sqrt{2}}, \quad V=\frac{z-t}{\sqrt{2}}.
\ea
In Brinkmann coordinates \cite{brinkmann}, the metric for a gravitational or electromagnetic $pp$-wave takes the form:
\begin{equation}\label{Brinkmann}
	g = 2dUdV + 2 H(U,X,Y)dU^2 + dX^2 + dY^2.
\end{equation}
For this spacetime geometry, the only non-zero curvature components are the Weyl and Ricci scalars, which reduce to:
\begin{equation}
	\psi_4 = \frac{1}{2}(H_{XX}-H_{YY}-2iH_{XY}), \quad
	\phi_{22} = \frac{1}{2}(H_{XX}+H_{YY}).
\end{equation}
Within $pp$-wave geometries \cite{ehlers-kundt,pirani}, the gravitational wave profile is governed entirely by a single complex Weyl component, $\psi_4=|\psi_4| e^{i\theta}$. This quantity varies only with the null parameter $U$, where its magnitude $|\psi_4|$ gives the wave's amplitude and the phase $\theta$ defines its polarization state. For instance, a fixed phase $\theta$ corresponds to a linearly polarized wave configuration. Conversely, the matter distribution embedded within the $pp$-wave background is described by the remaining non-vanishing Ricci component, $\phi_{22}$.

An extension of impulsive wavefronts is the sandwich wave profile. These configurations are mathematically modeled using Heaviside step functions that constrain the curvature fields to a restricted interval, $0 \leq U \leq U_0$. Consequently, the spacetime is flat and governed by the Minkowski metric both ahead of the incoming wave front ($U < 0$) and behind the passing tail ($U > U_0$). Within Einstein-Maxwell electrovacuum theory, a comprehensive derivation of sandwich plane wave metrics was obtained by \cite{halilsoy}. By adjusting the underlying boundary conditions, this formulation allows for the distinct study of pure gravitational radiation, pure electromagnetic fields or interacting electro-gravitational states that are confined to the localized curvature zone.

Plane waves constitute a highly symmetric subclass of general $pp$-waves characterized by spatial homogeneity across each individual wavefront. In terms of the Newman-Penrose parameters introduced above, this translation invariance implies that $\psi_4$ and $\phi_{22}$ must be independent of the transverse spatial coordinates $X$ and $Y$. For a linearly polarized sandwich plane wave, both the gravitational radiation field $\psi_4$ and the electromagnetic energy density $\phi_{22}$ exhibit no transverse coordinate dependence. Under these geometric constraints, the characteristic metric function simplifies to:
\begin{equation}
	H(U,X,Y)=\frac{1}{2}\big[\Theta(U)-\Theta(U-U_0)\big]\big[a^2(X^2+Y^2)-b^2(X^2-Y^2)\big],
\end{equation}
where the constant parameters $a$ and $b$ scale the respective contributions of the electromagnetic background and the gravitational wave amplitude.

To explicitly map out the transverse geometry of these spacetimes, it is frequently advantageous to transition to Rosen's metric formulation \cite{rosen}. This coordinate shift is achieved via the transformation \cite{Dereli_2020}:
\ba
U=u, \quad U_0=u_0, \nonumber \\ 
V=v + \frac{1}{2}(x^2FF_u+y^2GG_u),\nonumber \\ 
X=xF, \quad Y=yG
\ea
which recasts the line element into the diagonal form:
\begin{equation}\label{Rosen}
	g = 2dudv + F(u)^2 dx^2 + G(u)^2 dy^2.
\end{equation}
The structural functions $F(u)$ and $G(u)$ are determined by the decoupled system of second-order differential equations:
\ba
F_{uu} + A^2\big[\Theta(u)-\Theta(u-u_0)\big]F &= 0, \nonumber \\
G_{uu} + B^2\big[\Theta(u)-\Theta(u-u_0)\big]G &= 0.
\ea
By integrating across the boundary, the global solution can be written as:
\ba \label{eq9}
F(u) &= \cos\big[A\big(u\Theta(u)-(u-u_0)\Theta(u-u_0)\big)\big] - A\sin(Au_0)(u-u_0)\Theta(u-u_0), \nonumber \\
G(u) &= \cos\big[B\big(u\Theta(u)-(u-u_0)\Theta(u-u_0)\big)\big] - B\sin(Bu_0)(u-u_0)\Theta(u-u_0),
\ea
where the auxiliary constants are defined in terms of the physical parameters via $A^2 = a^2 - b^2$ and $B^2 = a^2 + b^2$.

To evaluate the exact radiation and matter profiles, we construct a localized null tetrad frame:
\begin{equation}
	l = du, \quad n = dv, \quad m = \frac{1}{\sqrt{2}}(Fdx+iGdy), \quad \bar{m} = \frac{1}{\sqrt{2}}(Fdx-iGdy).
\end{equation}
Projecting the curvature tensors onto this basis yields the following non-vanishing Newman-Penrose scalars:
\ba
\psi_4 &= -\frac{F_{uu}}{2F} + \frac{G_{uu}}{2G} = -b^2\big[\Theta(u)-\Theta(u-u_0)\big], \nonumber \\
\phi_{22} &= -\frac{F_{uu}}{2F} - \frac{G_{uu}}{2G} = a^2\big[\Theta(u)-\Theta(u-u_0)\big].
\ea
The remainder of this work investigates the behavior of the gravitational wave background across a pure gravitational field. The configuration can be extracted systematically as a limiting case of the general solution derived above. The structure of this sandwich wave spacetime is schematically outlined in Figure 1.

\clearpage

\begin{figure}[htb!]
	\centering
	\includegraphics[width=1.00 \textwidth]{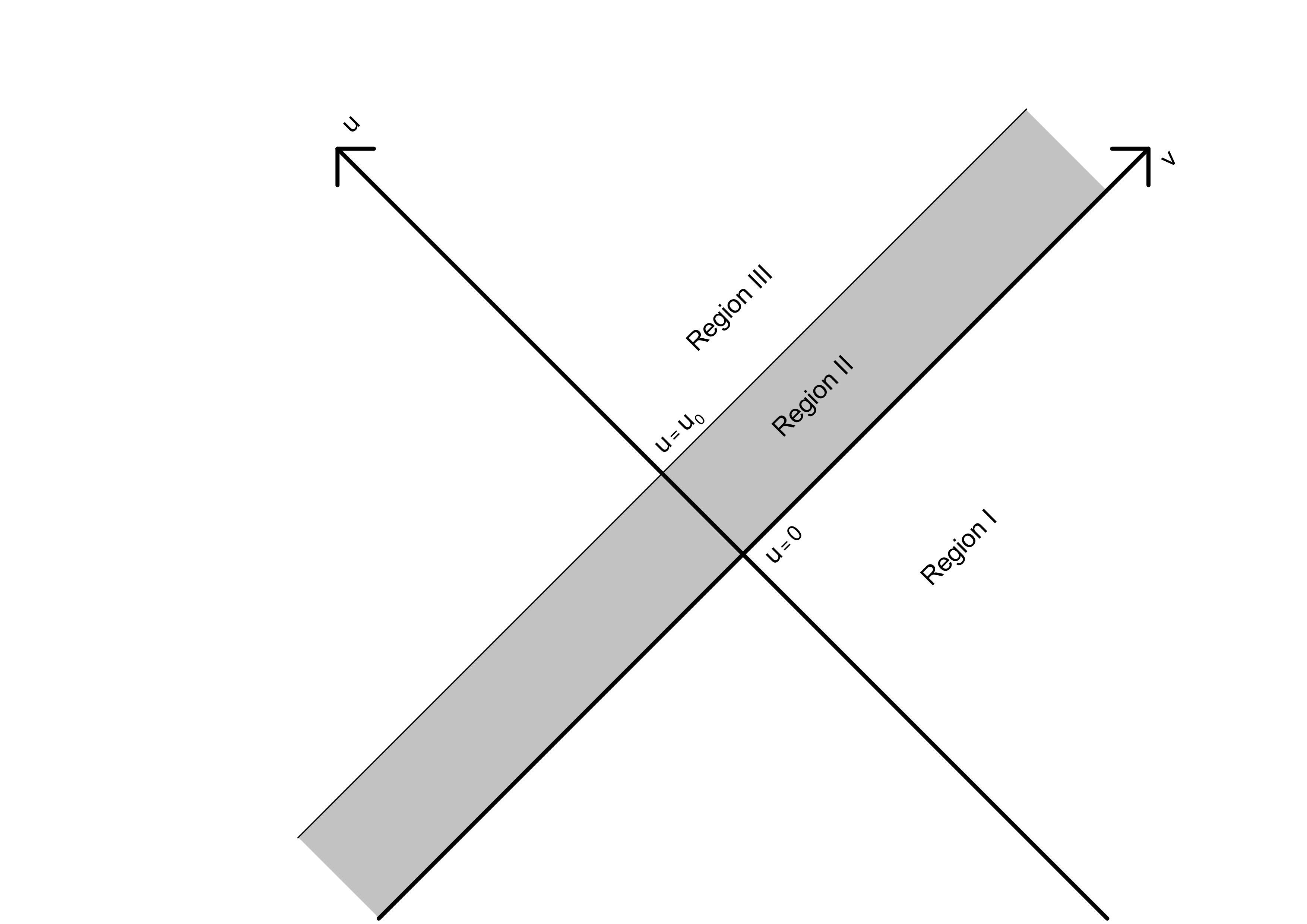}
	\caption{The general structure of the gravitational sandwich wave geometry consists of three regions. Region I represents the flat Minkowski spacetime ahead of the sandwich wave. Region II is the curved domain where the sandwich wave is present. Region III corresponds to the flat Minkowski region trailing the wave.}
\end{figure}

\noindent The three distinct regions, consisting of two flat regions on either side and a curved region within the finite duration plane fronted wave, must be smoothly joined using appropriate junction conditions. Previous studies \cite{halilsoy} have demonstrated that the O'Brien and Synge junction conditions are satisfied across the boundaries at $u=0$ and $u=u_0$ for the global sandwich wave solution in Einstein-Maxwell theory.

The metric functions characterizing a purely gravitational sandwich wave are derived by setting $a=0$ in (\ref{eq9}), resulting in:
\ba
F(u)&=&\left\{
\begin{array}{ll}
	1 , \quad u<0, & Region \hspace{2mm}I \\
	cosh(bu), \quad 0<u<u_0, & Region \hspace{2mm}I \\
	\alpha_0+\beta_0u, \quad u_0<u, & Region \hspace{2mm}III \\
\end{array}\right. \\
G(u)&=&\left\{
\begin{array}{ll}
	1 , \quad u<0, & Region \hspace{2mm}I \\
	cos(bu), \quad  0<u<u_0, & Region \hspace{2mm}II \\
	\gamma_0-\tau_0u, \quad u_0<u, & Region \hspace{2mm}III \\
\end{array}\right.
\ea
where
\ba
\alpha_{0}&=&cosh(bu_{0})-bu_{0}sinh(bu_{0}), \nonumber \\
\beta_{0}&=&bsinh(bu_{0}), \nonumber \\
\gamma_{0}&=&cos(bu_{0})+bu_{0}sin(bu_{0}), \nonumber \\
\tau_{0}&=&bsin(bu_{0}). \nonumber
\ea
To ensure that the coordinate patch remains completely regular and free of premature coordinate singularities within the interior wave zone, we naturally restrict the null duration of the curvature pulse to satisfy $u_0 < \frac{\pi}{2b}$. This bound guarantees that $G(u) = \cos(bu)$ remains strictly positive and non-vanishing throughout Region II ($0 < u < u_0$), ensuring that coordinate focusing effects occur only after the wave has passed.

\bigskip
\section{Neutrino Equation in a Gravitational Sandwich Plane Wave Spacetime}
A massive test Majorana neutrino field $\Phi$ is given by the 2-component spinor
\ba
\Phi = \left ( \begin{array}{c} \varphi_{1} \\ \varphi_{2} \end{array}  \right ),
\ea
satisfying the Weyl equation
\ba
\sigma^{a} \nabla_{X_{a}} \Phi =m \dot{\Phi},
\ea
where $\sigma^a :\{I ,\sigma^1, \sigma^2, \sigma^3 \}$ represent the Pauli matrices, while $\nabla_{X_{a}}$ denotes the covariant derivatives with respect to an orthonormal co-frame $\{e^a\}$. Therefore the covariant exterior derivative operator is given by $\nabla = e^a \nabla_{X_{a}}$.

$\Phi$ denotes a left-handed Weyl spinor, while $\dot{\Phi}$ denotes the corresponding dotted (right-handed) Weyl spinor transforming in the complex conjugate representation; the dot labels the spinor representation and does not indicate a derivative. The relation to complex conjugation involves the antisymmetric spinor metric (equivalently $\sigma_2$ in Pauli-matrix conventions), which is implicit in our notation.

\subsection{The Coupled Flavor Equations}
Let $\varphi_1, \varphi_2$ represent the two (odd-Grassmann) components of the electron neutrino flavor spinor ($\nu_e$), and let $\chi_1, \chi_2$ represent the (odd-Grassmann) components of the muon neutrino flavor spinor ($\nu_{\mu}$). All the components are  taken as complex valued functions of all coordinates $\{u,v,x,y\}$. 

To track the exact component behaviors, we define four spatial-temporal differential operators containing the metric functions $F(u)$ and $G(u)$ of the sandwich wave spacetime:
\begin{align*}
	D_1 &= \frac{\partial}{\partial v} \\
	D_2 &= \frac{1}{\sqrt{2}} \left( \frac{1}{F} \frac{\partial}{\partial x} - \frac{i}{G} \frac{\partial}{\partial y} \right) \\
	D_3 &= -\frac{\partial}{\partial u} - \frac{1}{2} \left( \frac{F_u}{F} + \frac{G_u}{G} \right) \\
	D_4 &= \frac{1}{\sqrt{2}} \left( \frac{1}{F} \frac{\partial}{\partial x} + \frac{i}{G} \frac{\partial}{\partial y} \right)
\end{align*}

The coupled Majorana equations for the two-flavor system are written explicitly as:
\begin{align}
	D_1 \varphi_1 + D_2 \varphi_2 &= \frac{1}{\sqrt{2}} \left( m_{ee} \varphi_2^* + m_{e\mu} \chi_2^* \right) \\
	D_3 \varphi_2 + D_4 \varphi_1 &= -\frac{1}{\sqrt{2}} \left( m_{ee} \varphi_1^* + m_{e\mu} \chi_1^* \right) \\
	D_1 \chi_1 + D_2 \chi_2 &= \frac{1}{\sqrt{2}} \left( m_{e\mu} \varphi_2^* + m_{\mu\mu} \chi_2^* \right) \\
	D_3 \chi_2 + D_4 \chi_1 &= -\frac{1}{\sqrt{2}} \left( m_{e\mu} \varphi_1^* + m_{\mu\mu} \chi_1^* \right)
\end{align}

Grouping these systems by their sharing operators yields two matrix differential systems:
\begin{align}
	D_1 \begin{pmatrix} \varphi_1 \\ \chi_1 \end{pmatrix} + D_2 \begin{pmatrix} \varphi_2 \\ \chi_2 \end{pmatrix} &= \frac{1}{\sqrt{2}} M_M \begin{pmatrix} \varphi_2^* \\ \chi_2^* \end{pmatrix} \\
	D_3 \begin{pmatrix} \varphi_2 \\ \chi_2 \end{pmatrix} + D_4 \begin{pmatrix} \varphi_1 \\ \chi_1 \end{pmatrix} &= -\frac{1}{\sqrt{2}} M_M \begin{pmatrix} \varphi_1^* \\ \chi_1^* \end{pmatrix}
\end{align}
where $M_M$ is the symmetric complex Majorana flavor mass matrix:
\[
M_M = \begin{pmatrix} m_{ee} & m_{e\mu} \\ m_{e\mu} & m_{\mu\mu} \end{pmatrix}
\]
\subsection{Unitary Transformation and Diagonalization}
We introduce the $2\times2$ PMNS\cite{Sakata,Pontecorvo,Pontecorvo2} mixing matrix $U$ parameterized by the mixing angle $\theta$ and the Majorana phase $\phi$:
\[
U = \begin{pmatrix} \cos\theta & \sin\theta e^{i\phi} \\ -\sin\theta e^{-i\phi} & \cos\theta \end{pmatrix}
\]
The mass eigenstates $\xi$ and $\eta$ are defined via the rotations:
\[
\begin{pmatrix} \varphi_1 \\ \chi_1 \end{pmatrix} = U \begin{pmatrix} \xi_1 \\ \eta_1 \end{pmatrix}, \quad \begin{pmatrix} \varphi_2 \\ \chi_2 \end{pmatrix} = U \begin{pmatrix} \xi_2 \\ \eta_2 \end{pmatrix}
\]
Taking the complex conjugate of these transformation yields:
\[
\begin{pmatrix} \varphi_1^* \\ \chi_1^* \end{pmatrix} = U^* \begin{pmatrix} \xi_1^* \\ \eta_1^* \end{pmatrix}, \quad \begin{pmatrix} \varphi_2^* \\ \chi_2^* \end{pmatrix} = U^* \begin{pmatrix} \xi_2^* \\ \eta_2^* \end{pmatrix}
\]
Substituting these field definitions into the matrix systems, and noting that the scalar operators $D_i$ commute with $U$, we multiply from the left, both sides by $U^\dagger$. Using $U^\dagger U = \mathbb{I}$, we obtain:
\begin{align}
	D_1 \begin{pmatrix} \xi_1 \\ \eta_1 \end{pmatrix} + D_2 \begin{pmatrix} \xi_2 \\ \eta_2 \end{pmatrix} &= \frac{1}{\sqrt{2}} \left( U^\dagger M_M U^* \right) \begin{pmatrix} \xi_2^* \\ \eta_2^* \end{pmatrix} \\
	D_3 \begin{pmatrix} \xi_2 \\ \eta_2 \end{pmatrix} + D_4 \begin{pmatrix} \xi_1 \\ \eta_1 \end{pmatrix} &= -\frac{1}{\sqrt{2}} \left( U^\dagger M_M U^* \right) \begin{pmatrix} \xi_1^* \\ \eta_1^* \end{pmatrix}
\end{align}
By Takagi factorization\cite{Takagi}, the combination $U^\dagger M_M U^* = M_{\text{diag}}$, gives the real physical mass eigenvalues:
\[
M_{\text{diag}} = \begin{pmatrix} m_1 & 0 \\ 0 & m_2 \end{pmatrix}
\]
where the explicit mass equations are:
\begin{align*}
	m_1 &= \frac{1}{2} \left[ (m_{ee} + m_{\mu\mu}) - \sqrt{(m_{ee} - m_{\mu\mu})^2 + 4m_{e\mu}^2} \right] \\
	m_2 &= \frac{1}{2} \left[ (m_{ee} + m_{\mu\mu}) + \sqrt{(m_{ee} - m_{\mu\mu})^2 + 4m_{e\mu}^2} \right]
\end{align*}

\subsection{The Completely Decoupled System}
Evaluating the diagonalized system breaks the equations into two completely isolated parts matching our original single-flavor formulation:

\subsubsection{Sector 1 (Mass Eigenstate 1: $\xi_1, \xi_2$)}
\begin{align}
	\left( \frac{\partial}{\partial v} \right) \xi_1 +  \frac{1}{\sqrt{2}} \left( \frac{1}{F} \frac{\partial}{\partial x}  - \frac{i}{G} \frac{\partial}{\partial y}  \right) \xi_2 &= \frac{m_1}{\sqrt{2}}\xi_2^* \\
	\left( -\frac{\partial}{\partial u}  -\frac{1}{2} \left( \frac{F_u}{F} +  \frac{G_u}{G} \right) \right) \xi_2 +  \frac{1}{\sqrt{2}} \left( \frac{1}{F} \frac{\partial}{\partial x} + \frac{i}{G} \frac{\partial}{\partial y}  \right) \xi_1 &= -\frac{m_1}{\sqrt{2}}\xi_1^*
\end{align}

\subsubsection{Sector 2 (Mass Eigenstate 2: $\eta_1, \eta_2$)}
\begin{align}
	\left( \frac{\partial}{\partial v} \right) \eta_1 +  \frac{1}{\sqrt{2}} \left( \frac{1}{F} \frac{\partial}{\partial x}  - \frac{i}{G} \frac{\partial}{\partial y}  \right) \eta_2 &= \frac{m_2}{\sqrt{2}}\eta_2^* \\
	\left( -\frac{\partial}{\partial u}  -\frac{1}{2} \left( \frac{F_u}{F} +  \frac{G_u}{G} \right) \right) \eta_2 +  \frac{1}{\sqrt{2}} \left( \frac{1}{F} \frac{\partial}{\partial x} + \frac{i}{G} \frac{\partial}{\partial y}  \right) \eta_1 &= -\frac{m_2}{\sqrt{2}}\eta_1^*
\end{align}

A class of exact solutions is expressed as
\begin{eqnarray}
	\xi_1 &=& \frac{i}{p_v\sqrt{2}\sqrt{FG}}\Bigg(  c_1\Big(\frac{ip_1}{F}+\frac{p_2}{G}\Big)-c_2^*m_1 \Bigg)e^{i(p_v v + p_1 x + p_2 y)}e^{iK_1(u)} \nonumber \\
	&+& \frac{i}{p_v\sqrt{2}\sqrt{FG}}\Bigg(  c_2\Big(\frac{ip_1}{F}+\frac{p_2}{G}\Big)+c_1^* m_1 \Bigg)e^{-i(p_v v + p_1 x + p_2 y)}e^{-iK_1(u)} \nonumber\\
	\xi_2 &=& \frac{c_1}{\sqrt{FG}} e^{i(p_v v + p_1 x + p_2 y)}e^{iK_1(u)}+\frac{c_2}{\sqrt{FG}} e^{-i(p_v v + p_1 x + p_2 y)}e^{-iK_1(u)}.
\end{eqnarray}
and
\begin{eqnarray}
	\eta_1 &=& \frac{i}{k_v\sqrt{2}\sqrt{FG}}\Bigg(  c_3\Big(\frac{ik_1}{F}+\frac{k_2}{G}\Big)-c_4^*m_2 \Bigg)e^{i(k_v v + k_1 x + k_2 y)}e^{iK_2(u)} \nonumber \\
	&+& \frac{i}{k_v\sqrt{2}\sqrt{FG}}\Bigg(  c_4\Big(\frac{ik_1}{F}+\frac{k_2}{G}\Big)+c_3^* m_2 \Bigg)e^{-i(k_v v + k_1 x + k_2 y)}e^{-iK_2(u)} \nonumber\\
	\eta_2 &=& \frac{c_3}{\sqrt{FG}} e^{i(k_v v + k_1 x + k_2 y)}e^{iK_2(u)}+\frac{c_4}{\sqrt{FG}} e^{-i(k_v v + k_1 x + k_2 y)}e^{-iK_2(u)}.
\end{eqnarray}
\noindent
where $p_v, p_1, p_2, k_v, k_1, k_2,$ represent constant momentum components, and $c_1,c_2,c_3,c_4$ are complex-valued odd-Grassmann parameters. The phase functions $K_1(u), K_2(u)$ are determined separately in each region by computing the relevant integral
\ba
K_1(u) &=& -\frac{1}{2 p_v} \int_{0}^{u} \left ( \frac{p_1^2}{F(u^{\prime})^{2}} + \frac{p_2^2}{G(u^{\prime})^{2}}+m_1^2  \right ) du^{\prime} \\ \nonumber 
K_2(u) &=& -\frac{1}{2 k_v} \int_{0}^{u} \left ( \frac{k_1^2}{F(u^{\prime})^{2}} + \frac{k_2^2}{G(u^{\prime})^{2}}+m_2^2  \right ) du^{\prime}.
\ea

	Now that the decoupled systems for $\xi_i$ and $\eta_i$ are resolved across the sandwich wave regions, the original electron and muon neutrino flavor states are reconstructed explicitly using the following linear combinations:
\begin{align}
	\varphi_1(u,v,x,y) &= \xi_1 \cos\theta + \eta_1 \sin\theta e^{i\phi} \\
	\varphi_2(u,v,x,y) &= \xi_2 \cos\theta + \eta_2 \sin\theta e^{i\phi} \\
	\chi_1(u,v,x,y) &= -\xi_1 \sin\theta e^{-i\phi} + \eta_1 \cos\theta \\
	\chi_2(u,v,x,y) &= -\xi_2 \sin\theta e^{-i\phi} + \eta_2 \cos\theta
\end{align}
\section{Analysis}
\subsection{Gravitational Modulation of Flavor Survival}

To isolate the explicit consequences of a localized spacetime curvature pulse on quantum flavor mixing, we evaluate a model where a pure electron neutrino beam is generated in the flat domain of Region I ($u < 0$). In this initial flat configuration, the metric functions are trivial ($F = G = 1$), and the system reduces to standard vacuum oscillations driven by the mass-squared difference and $\theta$.

We establish the incoming wave profile at the first boundary layer ($u = 0^-$) by setting the initial flavor amplitudes for the active component as:
\ba
\varphi_2(0, v, x, y) = \nu_0 e^{i(p_v v + p_1 x + p_2 y)}, \quad \chi_2(0, v, x, y) = 0
\ea
where $\nu_0$ scales the initial quantum amplitude of the electron flavor field. Projecting this flavor configuration onto the mass eigenstate basis at $u=0$ fixes the Grassmann constants in terms of the mixing angle:
\ba
c_1 = \nu_0 \cos\theta, \quad c_3 = \nu_0 \sin\theta e^{-i\phi}
\ea
When this propagating wavefront crosses $u = 0$, the transition to the anisotropic, polarized metric of Region $II$ ($0 \le u \le u_0$) fundamentally reshapes the evolution of the system. Rather than altering the underlying flavor mixing matrix itself, which remains fixed by the invariant parameters of the Takagi factorization, the spacetime background acts on the independent phase functions $K_1(u)$ and $K_2(u)$ of the decoupled mass fields $\xi$ and $\eta$.

To find the flavor evolution across the sandwich profile, we integrate the phase functions piecewise across each region. For the first mass eigenstate $\xi$, this yields the explicit expressions for $K_1(u)$:
\ba
K_1(u) = \left\{
\begin{array}{ll}
	-\dfrac{1}{2p_v}\left(p_1^2 + p_2^2 + m_1^2\right)u , \quad & Region \hspace{2mm}I \\ \\
	-\dfrac{1}{2p_v}\left[\dfrac{p_1^2}{b}\tanh(bu) + \dfrac{p_2^2}{b}\tan(bu) + m_1^2 u\right] , \quad & Region \hspace{2mm}II \\ \\
	-\dfrac{1}{2p_v}\left[ \dfrac{p_2^2}{\tau_0(\gamma_0-\tau_0u)} - \dfrac{p_1^2}{\beta_0(\alpha_0+\beta_0u)} + m_1^2 u + \Lambda_0 \right] , \quad & Region \hspace{2mm}III \\
\end{array}\right.
\ea
Similarly, the phase function $K_2(u)$ for the second mass eigenstate $\eta$ is given explicitly by:
\ba
K_2(u) = \left\{
\begin{array}{ll}
	-\dfrac{1}{2k_v}\left(k_1^2 + k_2^2 + m_2^2\right)u , \quad & Region \hspace{2mm}I \\ \\
	-\dfrac{1}{2k_v}\left[\dfrac{k_1^2}{b}\tanh(bu) + \dfrac{k_2^2}{b}\tan(bu) + m_2^2 u\right] , \quad & Region \hspace{2mm}II \\ \\
	-\dfrac{1}{2k_v}\left[ \dfrac{k_2^2}{\tau_0(\gamma_0-\tau_0u)} - \dfrac{k_1^2}{\beta_0(\alpha_0+\beta_0u)} + m_2^2 u + \Lambda_0' \right] , \quad & Region \hspace{2mm}III \\
\end{array}\right.
\ea
The integration constants $\Lambda_0$ and $\Lambda_0'$ are uniquely fixed by demanding that the mass-eigenstate wave functions remain globally continuous at $u = u_0$. Matching the boundary profiles yields:

\ba
\Lambda_0 &=& \frac{p_1^2}{b}\coth(bu_0) - \frac{p_2^2}{b}\cot(bu_0) \\
\Lambda_0' &=& \frac{k_1^2}{b}\coth(bu_0) - \frac{k_2^2}{b}\cot(bu_0)
\ea

To construct the physical transition probability $P_{\nu_e \to \nu_\mu}(u)$ in Region $III$, we evaluate the ratio of the muon-flavor field density to the total active state normalization factor:
\ba
P_{\nu_e \to \nu_\mu}(u) = \frac{|\chi_2(u)|^2}{\text{Norm}}
\ea
where the total normalization parameter $\text{Norm}$ is defined explicitly by summing the active components across the flavor basis:
\ba
\text{Norm} = |\varphi_2(u,v,x,y)|^2 + |\chi_2(u,v,x,y)|^2 = \frac{|c_1|^2 + |c_3|^2}{F(u)G(u)} = \frac{|\nu_0|^2}{F(u)G(u)}
\ea
Evaluating $|\chi_2(u)|^2$ using the explicit mass fields and substituting this normalization yields the expression:
\ba
P_{\nu_e \to \nu_\mu}(u) = \frac{1}{2} \sin^2(2\theta) \Big[ 1 - \cos\big(\Delta \Phi(u,v,x,y)\big) \Big]
\ea
where the complete phase difference driving the quantum flavor modulation is given by:
\ba
\Delta\Phi(u,v,x,y) = (p_v - k_v)v + (p_1 - k_1)x + (p_2 - k_2)y + \big(K_1(u) - K_2(u)\big)
\ea
Expanding the difference $K_1(u) - K_2(u)$ in the flat after-zone ($u > u_0$) allows us to decompose the phase structure into longitudinal, transverse, kinematic and memory parts:
\ba
\Delta\Phi(u,v,x,y) = \Delta\Phi_{\text{long}} + \Delta\Phi_{\text{trans}} + \Delta\Phi_{\text{kin}} + \Delta\Phi_{\text{mem}}
\ea
where the four individual contributions are defined explicitly as:
\ba
\Delta\Phi_{\text{long}} &\equiv& (p_v - k_v)v \\
\Delta\Phi_{\text{trans}} &\equiv& (p_1 - k_1)x + (p_2 - k_2)y \\
\Delta\Phi_{\text{kin}} &\equiv& \frac{1}{2}\left(\frac{m_2^2}{k_v} - \frac{m_1^2}{p_v}\right)u
\ea
\ba
\Delta\Phi_{\text{mem}} &\equiv& \frac{1}{2k_v}\left[\frac{k_2^2}{\tau_0(\gamma_0-\tau_0u)} - \frac{k_1^2}{\beta_0(\alpha_0+\beta_0u)} + \Lambda_0'\right] \nonumber \\
&-& \frac{1}{2p_v}\left[\frac{p_2^2}{\tau_0(\gamma_0-\tau_0u)} - \frac{p_1^2}{\beta_0(\alpha_0+\beta_0u)} + \Lambda_0\right]
\ea
The explicit mathematical structure of $P_{\nu_e \to \nu_\mu}(u)$ uncovers two important features that diverge from standard Minkowski space studies. First, because the gravitational wave breaks the kinematic alignment of the mass states ($p_\mu \neq k_\mu$), the transverse variables $(x,y)$ and null variable $v$ remain explicitly within $\Delta\Phi_{\text{long}}$ and $\Delta\Phi_{\text{trans}}$. Consequently, the neutrino beam develops an four-dimensional interference grid. Rather than oscillating uniformly as a function of distance, the instantaneous flavor ratio varies locally across the transverse profile of the wavefront, preserving a permanent phase memory effect $\Delta\Phi_{\text{mem}}$ of the curvature profile.

\subsection{Gravitationally Induced Helicity Flips}

While the index-2 spinor components govern the dynamic flavor-modulation profile of the neutrino beam, the accompanying index-1 components ($\varphi_1, \chi_1$) track the evolution of the system's underlying helicity states. In a massive Majorana field framework, the localized spacetime curvature pulse lacks spherical symmetry, allowing the gravitational background to couple directly to the spin-tensor structure. This interaction drives transitions between active left-handed and right-handed configurations. 

To present this mechanism of helicity-flipping in the flat after-zone ($u > u_0$), we define the time-dependent transverse geometric momentum operators for each respective mass eigenstate path:
\ba \label{geom_mom}
\mathcal{V}_1(u) \equiv \frac{i p_1}{F(u)} + \frac{p_2}{G(u)} = \frac{i p_1}{\alpha_0+\beta_0u} + \frac{p_2}{\gamma_0-\tau_0u} \\
\mathcal{V}_2(u) \equiv \frac{i k_1}{F(u)} + \frac{k_2}{G(u)} = \frac{i k_1}{\alpha_0+\beta_0u} + \frac{k_2}{\gamma_0-\tau_0u}
\ea
The total probability of a gravitationally induced helicity transition, $P_{\text{flip}}(u,v,x,y)$, is evaluated by calculating the ratio of the total field density to the normalization parameter:
\ba \label{p_flip_def}
P_{\text{flip}}(u,v,x,y) = \frac{|\varphi_1(u)|^2 + |\chi_1(u)|^2}{\text{Norm}}
\ea
where the normalization factor $\text{Norm}$ incorporates the transverse expansion:
\ba \label{norm_def}
\text{Norm} \equiv |\varphi_2(u,v,x,y)|^2 + |\chi_2(u,v,x,y)|^2 = \frac{|\nu_0|^2}{F(u)G(u)}
\ea
Substituting the independent mass-eigenstate fields into the flavor equations and carrying out the algebra yields the explicit analytical expression for the net spin-flip probability within Region $III$:
\ba \label{p_flip_final}
P_{\text{flip}}(u,v,x,y) = \frac{1}{2}\left[ \frac{\cos^2\theta}{p_v^2}|\mathcal{V}_1(u)|^2 + \frac{\sin^2\theta}{k_v^2}|\mathcal{V}_2(u)|^2 \right] \nonumber \\
+ \frac{\sin^2(2\theta)}{4 p_v k_v} \left( \frac{p_v}{k_v} + \frac{k_v}{p_v} - 1 \right) \text{Re}\left( \mathcal{V}_1(u)\mathcal{V}_2^*(u) e^{i\Delta\Phi(u,v,x,y)} \right)
\ea
where the phase argument $\Delta\Phi(u,v,x,y)$ corresponds to the interference pattern defined in the previous subsection.

Equation (\ref{p_flip_final}) shows that the transition into sterile states is driven entirely by the alignment of the split momentum trajectories with the polarization axes of the metric functions. In flat Minkowski space where the drift velocities vanish ($\beta_0 = \tau_0 = 0$), these geometric functions reduce to constants proportional to the standard mass-energy ratio ($\sim m^2/E^2$), suppressing the helicity flip.

Behind the sandwich wave, the linear parameters $\beta_0$ and $\tau_0$ source the cross-coupling terms. The gravitational wave thus acts as a source of chiral mixing: as the neutrino propagates through Region $III$, active left-handed states are continuously converted into right-handed fields. This conversion depends on the transverse coordinates $(x,y)$, leaving a macroscopic signature of the gravitational wave in the helicity profile.

\subsection{Transverse Momentum Dependence and Wave Packet Separation}

The last feature arising from the interaction between the massive Majorana field and the sandwich wave background is the breaking of spatial symmetry across the transverse coordinate plane. In standard Minkowski space models, the choice of orientation within the transverse $(x,y)$-plane is physically arbitrary, as the background geometry is completely isotropic. However, the metric functions $F(u)$ and $G(u)$ introduce a polarization in Regions $II$ and $III$ that splits the propagation along separate coordinate axes. 

To isolate this transverse anisotropy explicitly, we examine the evolution of the geometric transverse velocity operators $\mathcal{V}_1(u)$ and $\mathcal{V}_2(u)$ derived in the previous subsection. By finding the explicit components from Eq. (\ref{geom_mom}), we obtain:
\ba
\text{Re}\big(\mathcal{V}_1(u)\big) = \frac{p_2}{\gamma_0-\tau_0u}, \quad \text{Im}\big(\mathcal{V}_1(u)\big) = \frac{p_1}{\alpha_0+\beta_0u}
\ea
This decomposition shows a divergence in how the background geometry affects the momentum components. Along the $x$-axis, the wave packet behavior depends on the drift parameters $\alpha_0$ and $\beta_0$. Because $\beta_0 = b\sinh(bu_0) > 0$, the denominator $(\alpha_0 + \beta_0 u)$ grows monotonically as the neutrino packet moves into the flat after-zone ($u > u_0$). This stretching causes the effective momentum contribution along the $x$-axis to undergo a geometric spreading:
\ba
\lim_{u \to \infty} \text{Im}\big(\mathcal{V}_1(u)\big) = 0
\ea
Conversely, the dynamics along the $y$-axis are depending on the drift parameters $\gamma_0$ and $\tau_0$. Because the interior solution in Region $II$ varies as $\cos(bu)$, the matching boundary coefficients scale as $\tau_0 = b\sin(bu_0)$. Consequently, the linear profile $(\gamma_0 - \tau_0 u)$ decreases monotonically as a function of the null coordinate $u$. This coordinate contraction focuses the beam, amplifying the spatial momentum along the $y$-axis:
\ba \label{u_focus}
\lim_{u \to u_{\text{focus}}} \left| \text{Re}\big(\mathcal{V}_1(u)\big) \right| = \infty, \quad \text{where} \quad u_{\text{focus}} = \frac{\gamma_0}{\tau_0} = \frac{1}{b}\cot(bu_0) + u_0
\ea
This asymmetry changes the physical observables $P_{\nu_e \to \nu_\mu}(u)$ and $P_{\text{flip}}(u)$. The flavor oscillations depend on the difference between the mass state phase functions $K_1(u) - K_2(u)$. Separating the transverse momentum contribution in the memory phase $\Delta\Phi_{\text{mem}}$ gives:
\ba \label{delta_phi_asym_x}
\Delta\Phi_{\text{mem}}^{(x)} = \frac{1}{2}\left( \frac{k_1^2}{k_v} - \frac{p_1^2}{p_v} \right) \left[ \frac{1}{\beta_0(\alpha_0+\beta_0u)} \right] 
\ea
\ba \label{delta_phi_asym_y}
\Delta\Phi_{\text{mem}}^{(y)} = \frac{1}{2}\left( \frac{k_2^2}{k_v} - \frac{p_2^2}{p_v} \right) \left[ \frac{1}{\tau_0(\gamma_0-\tau_0u)} \right]
\ea
The differences between Eq. (\ref{delta_phi_asym_x}) and Eq. (\ref{delta_phi_asym_y}) confirm that the gravitational wave induces an anisotropic dispersion. For a neutrino beam with an initial momentum vector aligned strictly along the $x$-axis ($p_2 = k_2 = 0$), the phase becomes stationary and the spatial oscillations cease over large distances. Conversely, for a beam aligned along the $y$-axis ($p_1 = k_1 = 0$) the spatial oscillations accelerate as the wavefront approaches the focal point $u_{\text{focus}}$ defined in Eq. (\ref{u_focus}).

Furthermore, this spatial anisotropy alters the production rate. The cross-interference term in the helicity transition probability $P_{\text{flip}}$ found in Eq. (\ref{p_flip_final}) depends explicitly on the real part of $\mathcal{V}_1(u)\mathcal{V}_2^*(u)$, which is:
\ba \label{cross_prod_v}
\text{Re}\left( \mathcal{V}_1(u)\mathcal{V}_2^*(u) \right) = \frac{p_1 k_1}{(\alpha_0+\beta_0u)^2} + \frac{p_2 k_2}{(\gamma_0-\tau_0u)^2}
\ea
According to Eq. (\ref{cross_prod_v}), the conversion rate is directionally selective. The metric expansion suppresses the right-handed state production driven by $x$-axis momentum, whereas the contraction along the $y$-axis enhances the spin-flip operator. The passing sandwich wave thus induces an asymmetry in the neutrino beam, splitting its flavor and helicity profiles along the polarization axes of the original gravitational pulse.
\section{Conclusion}

In this work, we have presented an analysis of flavor-mixing dynamics and helicity transitions for massive Majorana neutrinos propagating through an exact plane gravitational wave background. By embedding the neutrino field equations within the anisotropic geometry of a sandwich wave spanning three distinct spacetime domains, we found the exact evolution of the quantum states across the boundaries. Using the properties of the Takagi factorization, we decoupled the underlying mass eigenstates, reducing a complex coupled system of differential equations into independent, integrals of the phase.

The results in Section 4 show significant modifications to Minkowski predictions, arising from the gravitational background geometry. Specifically, the passing wave packet breaks the kinematic alignment of neutrino mass eigenstates ($p_\mu \neq k_\mu$), inducing complex quantum interference. Consequently, the flavor ratio varies across the transverse profile of the wavefront, creating a phase memory effect, $\Delta\Phi_{\text{mem}}$, that remains after the gravitational pulse passes.

In addition, the index-1 spinor components lead to chiral polarization. In the massive Majorana case, the broken spherical symmetry of the background couples to the spin-tensor structure, inducing left to right-handed state transitions. Behind the wave pulse, the velocity memory effect, parameterized by $\beta_0$ and $\tau_0$, describes the spin-flip evolution. This transition probability depends strongly on the transverse momentum direction: it is suppressed by the metric expansion along the $x$-axis and enhanced by the contraction along the $y$-axis.

In conclusion, these findings establish that a gravitational wave pulse leaves a coordinate-dependent signature on both the flavor and helicity profiles of a neutrino beam. This permanent modification provides a physical manifestation of the gravitational velocity memory effect, mapping the polarization axes of the exact plane wave onto measurable quantum observables. This framework could be applied to high-energy astrophysical environments, such as core collapse supernovae or primordial gravitational wave backgrounds, where the interaction between intense neutrino fluxes and localized spacetime fluctuations may lead to observable effects driven by gravitationally mediated conversions.
\bigskip

\section{Acknowledgements}
This article is dedicated to Prof. Tekin Dereli, with deepest gratitude for his invaluable guidance, inspiring mentorship, and profound dedication to mathematical physics.

\section*{Data Availability Statement}
No new data were created or analyzed in this study.

\end{document}